\documentclass[a4paper,11pt]{article}
\pdfoutput=1 

\usepackage{jcappub} 

\usepackage[T1]{fontenc} 
\usepackage[utf8]{inputenc}
\usepackage{babel}

\title{\boldmath Analytical warm dark matter power spectrum on small scales}

\author[a,b]{G. Pordeus-da-Silva}
\author[c]{R. C. Batista}
\author[c]{L. G. Medeiros}

\affiliation[a]{Escola CAIC José Joffily, Secretaria da Educação e da Ciência e Tecnologia, Governo da Paraíba, PB, Brasil.}
\affiliation[b]{Departamento de Física Teórica e Experimental - Universidade Federal do Rio Grande do Norte, Natal, RN, Brasil.}
\affiliation[c]{Escola de Ciências e Tecnologia, Universidade Federal do Rio Grande do
Norte, RN, Brasil}

\emailAdd{givalpordeus@ufrn.edu.br}
\emailAdd{rbatista@ect.ufrn.br}
\emailAdd{leo.medeiros@ufrn.br}

\abstract{Using the Reduced Relativistic Gas (RRG) model, we analytically determine
the matter power spectrum for Warm Dark Matter (WDM) on small scales,
$k>1\ h\text{/Mpc}$. The RRG is a simplified model for the ideal relativistic
gas, but very accurate in the cosmological context. In another work,
we have shown that, for typical allowed masses for dark matter particles,
$m>5\ \text{keV}$, the higher order multipoles, $\ell\ge2$, in the Einstein-Boltzmann
system of equations are negligible on scales $k<10\ h\text{/Mpc}$.
Hence, we can follow the perturbations of WDM using the ideal fluid
framework, with equation of state and sound speed of perturbations
given by the RRG model. We derive a Mészáros-like equation for WDM
and solve it analytically in radiation, matter and dark energy dominated
eras. Joining these solutions, we get an expression that determines
the value of WDM perturbations as a function of redshift and wavenumber.
Then we construct the matter power spectrum and transfer function
of WDM on small scales and compare it to some results coming from
Lyman-$\alpha$ forest observations. Besides being a clear and pedagogical
analytical development to understand the evolution of WDM perturbations,
our power spectrum results are consistent with the observations considered
and the other determinations of the degree of warmness of dark matter
particles.}

\begin{document}
\maketitle
\flushbottom

\section{Introduction}

Cold Dark Matter (CDM) is a key concept for the understanding
of the universe. Together with the Cosmological Constant, $\Lambda$,
the $\Lambda$CDM model provides a very good description of the universe
on large-scales, e.g., \cite{Abbott2021,Aghanim:2018eyx}. However,
on small scales ($5\ h/\text{Mpc}\lesssim k\lesssim50\ h/\text{Mpc}$),
the $\Lambda$CDM model is challenged by some inconsistencies between
predictions of structure formation simulations and observations, such
as the Cusp/Core, Too Big to Fail and Missing Satellites problems,
\cite{DelPopolo:2016emo,Kameli:2019bki,Perivolaropoulos2021}. Although
these problems can be related to complex astrophysical effects and/or
systematic errors, Warm Dark Matter (WDM) was proposed as a solution
to them \cite{Bode_2001,Viel2013,Adhikari_2017}, though see \cite{Schneider2014}.
Nevertheless, one can always ask ``How cold is dark matter?'' \cite{RRG2009Fabris,Armendariz-Picon:2013jej,Piattella:2015nda,Martins2018,Maccio2010}.

In this paper, we will analytically study the WDM effects on small
scales using the Reduced Relativistic Gas (RRG) model, which assumes
that all particles of a classical ideal gas have the same momentum
magnitude. Although this is a very strong simplifying assumption,
it was shown that, in the cosmological context, the model is very
accurate when compared with the usual relativistic Maxwell distribution
\cite{RRG2005BERREDO-PEIXOTO,RRG2009Fabris}. The RRG model
was used in several cosmological studies, e.g., \cite{RRG2005BERREDO-PEIXOTO,RRG2009Fabris,RRG2012Fabris,RRG2012Leo,RRG2014Fabris,RRG2018dosReis,RRG2018Hipolito,Abdalla2019,AgudeloRuiz2020}.
The main idea in the RRG model was also developed in \cite{Mastache2019,Mastache2020,Macorra2020},
however without recognizing the previous works about it.

In particular, we are interested in analyzing the effects of WDM on
the linear matter power spectrum on scales $k>1\ h\text{/Mpc}$.
This type of study is usually conducted with the use Einstein-Boltzmann
solvers, like CAMB \cite{Lewis2000} and CLASS \cite{Lesgourgues2011}, or intricate analytical \cite{deVega:2011gg} studies.
We will show that the RRG model provides a simpler, reliable and pedagogical framework to understand WDM perturbations on small scales. In our approach, it is possible to obtain analytical solutions of WDM first-order perturbations and compute the matter power spectrum with a few
more considerations about the effects of baryons and dark energy.

As discussed in \cite{Pordeus-da-Silva:2019bak}, for the typical
mass scales allowed for WDM, the Einstein-Boltzmann system of equations
for RRG is effectively described by a perfect fluid on scales $k<10\ h\text{/Mpc}$.
Hence, we can determine the evolution of first-order perturbations
of WDM using the RRG model in the perfect fluid approximation. Then we
derive and analytically solve a Mészáros-like equation for WDM. With
these solutions, we determine the power spectrum of WDM perturbations
for scales $k>1\ h\text{/Mpc}$, which, given the allowed range
of dark matter mass, are the scales most impacted by WDM.

Finally, we make use of this solution to compute the matter power
spectrum and transfer function of WDM. Using matter power spectrum
data derived from Lyman-$\alpha$ observations, compiled in Ref.~\cite{Chabanier:2019eai},
we make a simplified statistical analysis of the warmness parameter
of RRG to check the consistency of our results with known limits for
dark matter velocity dispersion \cite{Armendariz-Picon:2013jej}.
Although this is not a complete and consistent statistical exploration
of all relevant cosmological parameters, it can provide reasonable
velocity dispersion limits for WDM, showing that important effects
of non-cold dark matter particles are well captured by our analysis.

The plan for this paper is the following. In Sect.~\ref{Sect2} we review the
main ideas of the RRG model, including background and first-order perturbations
equations. In Sect.~\ref{Sect3}, we present the approximations used for the analysis
and derive the Mészáros equation for WDM. In Sect.~\ref{Sect4}, we solve the
Mészáros equation for WDM. In Sect.~\ref{Sect5}, we correct the pure WDM solutions
for the presence of baryons and dark energy. Finally, in Sect.~\ref{Sect6},
we construct the matter power spectrum, transfer function and make
a simplified statistical analysis of the warmness parameter of the
model in light of some of the available power spectrum data on small
scales obtained from Lyman-$\alpha$ forest observations.

\section{Review of RRG model \label{Sect2}}

\subsection{Background evolution}

In order to explore the cosmological evolution of WDM perturbations,
we will make use of the RRG model. Let us review the relevant
features of this model, presented and studied in \cite{RRG2005BERREDO-PEIXOTO,RRG2009Fabris,RRG2012Fabris,RRG2018dosReis,RRG2018Hipolito,Pordeus-da-Silva:2019bak}. The basic assumption in the RRG model is that all classical particles
that constitute an ideal gas have the same momentum magnitude. Then,
making use of basic kinetic theory, one finds the equation of state
of RRG 
\begin{equation}
P=\frac{\rho}{3}v_{\text{th}}^{2}=\frac{\rho}{3}\left[1-\left(\frac{\rho_{d}}{\rho}\right)^{2}\right]\text{ ,}\label{EoS RRG}
\end{equation}
where $P$, $\rho$, $v_{\text{th}}$, $\rho_{d}\equiv nm$ and $n$
are, respectively, pressure, energy density, thermal velocity, rest
energy density and number density of the gas.

Using the conservation law, $dU=-PdV$, one can determine the evolution
of RRG energy density as a function of the scale factor in an FLRW metric
\begin{equation}
\rho\left(a\right)=\rho_{d,\text{ref}}\left(\frac{a_{\text{ref}}}{a}\right)^{3}\sqrt{1+b^{2}\left(\frac{a_{\text{ref}}}{a}\right)^{2}}\ ,\label{eq Rho RRG fuc de a e b1}
\end{equation}
where $b$ is the dimensionless warmness parameter and $a_{\text{ref}}$
can be chosen arbitrarily. In what follows define $a_{\text{ref}}=1$,
so the gas is in the ultra-relativistic regime when $a\ll b$, whereas
for $a\gg b$ the gas is in the non-relativistic regime. From (\ref{EoS RRG}),
the Equation of State (EoS) parameter is given by:
\begin{equation}
w=\frac{1}{3}v_{\text{th}}^{2}=\frac{1}{3}\left[\frac{b^{2}}{a^{2}+b^{2}}\right]\,.
\label{eq:EoS}
\end{equation}

It is possible to relate the parameter $b$ with the mass of the particles that constitute the gas. Assuming an instantaneous transition from
ultra-relativistic to non-relativistic regimes occurs at $T\simeq m$
and that the gas was is in thermal equilibrium with photons in the early
universe, one finds 
\begin{equation}
m\simeq\sqrt{3}\left(\frac{T_{\gamma0}}{b}\right)\simeq\frac{4.07\times10^{-7}}{b}\text{ keV ,}\label{massa b WDM}
\end{equation}
see Ref.~\cite{Pordeus-da-Silva:2019bak} for details and Ref.~\cite{RRG2018Hipolito}
for another expression fitted using the RRG transfer function, which gives
similar results. Considering the current limits on WDM mass obtained
from Lyman-$\alpha$ forest observations, $m>5.3\ \text{keV}$ (2$\sigma$
C.L.) \cite{Irs2017}, typical values of the warmness parameter are
$b<10^{-7}$.

We will study WDM linear perturbations from radiation dominated era
until today, so the Hubble function is given by:
\begin{equation}
H^{2}=H_{0}^{2}\left[\Omega_{r}a^{-4}+\Omega_{m}a^{-3}\sqrt{1+b^{2}/a^{2}}+\Omega_{\Lambda}\right]\,,\label{eq:Hubble-func}
\end{equation}
where we assume $\Omega_{r}=\Omega_{\gamma}+\Omega_{\nu}=2.469\times10^{-5}\left(1+0.2271N_{{\rm eff}}\right)h^{-2}$,
the density parameter of ultra-relativistic components (photons plus
massless neutrinos) with $N_{{\rm eff}}=3.046$, $\Omega_{m}=\Omega_{dm}+\Omega_{b}$
is the matter density (baryons plus dark matter) and $\Omega_{\Lambda}=1-\Omega_{m}-\Omega_{r}$
is the Cosmological Constant density parameter (assuming flat spatial
section). Note that we treat DM and baryons as the same fluid. Given
the typical mass limits for WDM, this assumption has essentially no
effect on the background evolution.

\begin{figure}[!htb]
\centering{}\includegraphics[scale=0.6]{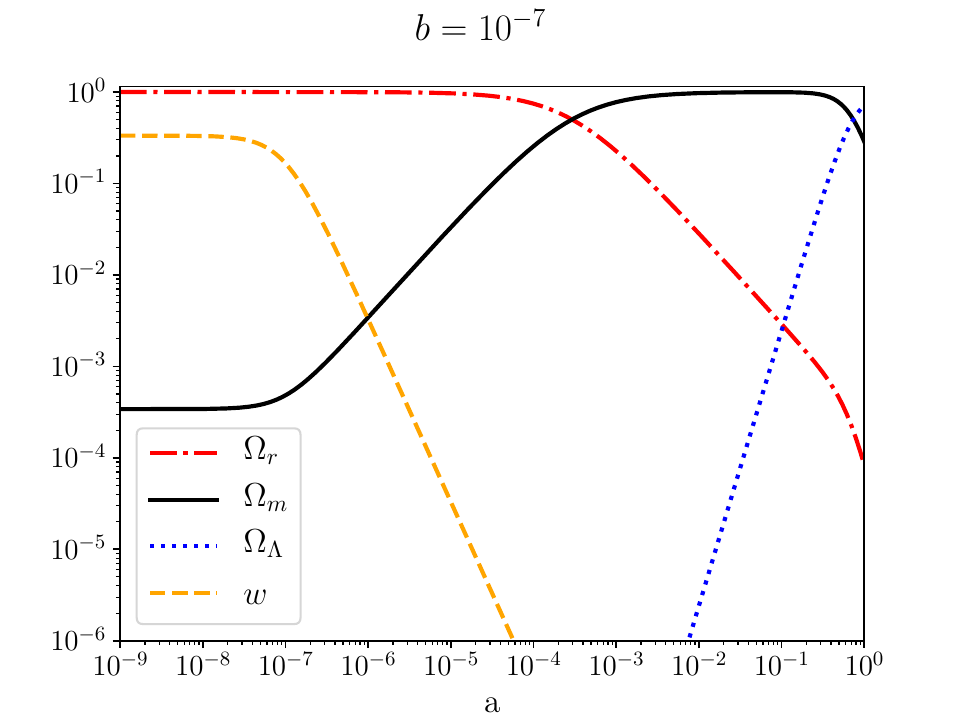}\caption{Evolution of background quantities for WDM with $b=10^{-7}$: $\Omega_{r}\left(a\right)$
(radiation, red dot-dashed line) $\Omega_{m}\left(a\right)$ (matter,
solid black line), $\Omega_{\Lambda}\left(a\right)$ (cosmological
constant, blue dotted line) and $w\left(a\right)$ (EoS parameter
of matter, dashed yellow line). Note that, for the chosen value of
$b$, $w\simeq1/3$ initially, but it rapidly decays several orders
of magnitude, indicating that, at low-$z$ dark matter is effectively
cold. It's also important to see that, despite $w\simeq1/3$ initially,
$\Omega_{m}$ is very small deep in the radiation dominated era. In
this plot we used $\Omega_{dm}=0.112h^{-2}$, $\Omega_{b}=0.02260h^{-2}$
and $h=0.6688$.\label{fig:background}}
\end{figure}

Although RRG with $b\sim10^{-7}$ is in the UR regime around the neutrino
decoupling time, $a\sim10^{-10}$, its contribution for the total
energy density of the universe at this epoch is very small, $\Omega_{m}\sim10^{-4}$,
see Figure~\ref{fig:background}. In this figure, we can also see that
$w$ decays rapidly during the radiation dominated era. Before the
matter-radiation equality, the initially hot dark matter is much cooler.
In this scenario, the background evolution is indeed quite close to
the $\Lambda$CDM one. As an example, the scale factor of radiation-matter
equality in a universe with WDM, $a_{{\rm eq}}$, is virtually the
result as in CDM:
\begin{equation}
a_{\text{eq}}=\sqrt{\left(\frac{\Omega_{r}}{\Omega_{m}}\right)^{2}-b^{2}}=\frac{\Omega_{r}}{\Omega_{m}}\left[1-\frac{1}{2}\left(\frac{b}{\Omega_{r}/\Omega_{m}}\right)^{2}+\mathcal{O}\left(\left(\frac{b}{\Omega_{r}/\Omega_{m}}\right)^{4}\right)\right]\text{\,.}\label{eq:a-equality}
\end{equation}
For $b=10^{-7}$, the first correction term is $\sim10^{-7}$ and
will be neglected.

\subsection{First-order perturbations}

One can also determine the properties and cosmological equations for
RRG making use of distribution functions \cite{RRG2018dosReis,Pordeus-da-Silva:2019bak},
both in the background and perturbative levels. The natural implementation
is 
\begin{equation}
f\left(p,\bar{p}\right)\equiv\frac{2\pi^{2}n}{\bar{p}^{3}}p\delta_{{\rm D}}\left(p-\bar{p}\right)\,,\label{eq:distribution-function}
\end{equation}
where $\delta_{{\rm D}}$ is the Dirac delta function and $\bar{p}$
is the momentum magnitude that all particles share. The advantage of
this description is that the Einstein-Boltzmann system of equations can
be derived for first-order RRG perturbations, which can be analyzed
in order to determine the circumstances under which the usual perfect
fluid approach is valid \cite{Pordeus-da-Silva:2019bak}.

In particular, we have previously shown in Ref.~\cite{Pordeus-da-Silva:2019bak} 
that, assuming the warm approximation, which considers
the first correction to the pressureless fluid, multipoles higher
than $\ell=2$ are strongly suppressed on scales $c_{s}k\eta\ll1$,
where the sound speed in rest frame is given by
\begin{equation}
c_{s}^{2}=\frac{\delta p}{\delta \rho}=\frac{w}{3}\left(\frac{5-3w}{1+w}\right)\,,
\label{eq:sound-speed}
\end{equation}
where $\delta \rho$ is the density perturbation and $\delta p$ the pressure perturbation of the RRG fluid. 
For typical allowed values of $b<10^{-7}$, the perfect fluid description
of warm RRG is valid for $k<10^{3}\ h \text{/Mpc}$ at $z=0$ and
$k<10\ h \text{/Mpc}$ at $z_{{\rm eq}}\sim10^{4}$. Therefore, when
studying WDM in the mass range $m>5\ \text{keV}$ or $b<10^{-7}$, the
perfect fluid description is well motivated. It was also shown that
only for masses above $400\ \text{keV}$, the gas is not relativistic
around the neutrino decoupling temperature, $T\simeq 1\ \text{MeV}$,
or $a\sim10^{-10}$. Hence, if dark matter is constituted by particles
with $m<400\ \text{keV}$, they can not be considered cold in the early
universe.

We also make some simplifying assumptions for the perturbations in
the relativistic components. We neglect $\ell\ge2$ multipoles
for photons and massless neutrinos, which then are considered the
same fluid, $\rho_{r}=\rho_{\gamma}+\rho_{\nu}$. For photons this
is valid while they are strongly coupled to baryons. After decoupling,
higher multipoles of photon perturbations become important, but,
since $\rho_{\gamma}\delta_{\gamma}$ is very subdominant with respect
to $\rho_{m}\delta_{m}$ after decoupling, the impact of higher photon
multipoles on matter perturbations is very small.

We are going to work in the Newtonian gauge, which, under the simplifying assumptions just described, have the following line element
\begin{equation}
ds^2 = a^2{\left( \eta \right)} \left[
-(1-2\Phi)d\eta ^2 + (1+2\Phi)d\vec{x}^2 \right]\,.
\label{metric}
\end{equation}
The ``Poisson'' equation (00 component of Einstein equations) is given by:
\begin{equation}
3\mathcal{H}\Phi^{\prime}+3\mathcal{H}^{2}\Phi+k^{2}\Phi=4\pi Ga^{2}\left(\rho_{m}\delta_{m}+\rho_{r}\delta_{r}\right)\,,\label{eq:poisson-full}
\end{equation}
where the prime represent time derivative with respect to the conformal
time, $\eta$, $\mathcal{H}=aH$ and $\delta_i=\delta \rho_i / \rho_i$ is the density contrast of matter ($i=m$) and ultra-relativistic components ($i=r$). Bear in mind that, initially,
we consider that $\rho_{m}\delta_{m}$ represent dark matter and
baryons contributions. Of course baryons are initially coupled to
photons and can not follow dark matter perturbations, so we will make
corrections to this effect later on.

We also need the dynamical equation for the potential ($ii$ component of Einstein equations), 
\begin{equation}
\Phi^{\prime\prime}+3\mathcal{H}\Phi^{\prime}+\left(2\frac{a^{\prime\prime}}{a}-\mathcal{H}^{2}\right)\Phi=4\pi Ga^{2}\left(c_{s}^{2}\rho_{m}\delta_{m}+\frac{1}{3}\rho_{r}\delta_{r}\right)\,.\label{eq:ddphi}
\end{equation}
The conservation equations for WDM perturbations are: 
\begin{equation}
\delta_{m}{}^{\prime}+3\mathcal{H}\left(c_{s}^{2}-w\right)\delta_{m}+\left(1+w\right)\left(kV_{m}+3\Phi{}^{\prime}\right)=0
\label{eq:cont-delta-m}
\end{equation}
and
\begin{equation}
V_{m}{}^{\prime}+\mathcal{H}\left(1-3c_{s}^{2}\right)V_{m}-k\left(\frac{c_{s}^{2}}{1+w}\delta_{m}-\Phi\right)=0\text{ },\label{eq:euler-delta-m}
\end{equation}
where $V_m$ is the peculiar velocity of WDM, as defined as in \cite{Piattella:2018}. Note that the sound speed in the rest frame is equal to adiabatic sound speed, i.e., $c_{s}^{2}=c_{a}^{2}=\dot{p}/\dot{\rho}$. Hence, the RRG
has no intrinsic non-adiabatic perturbations.

We can combine these equations to obtain a second-order differential
equation for $\delta_{m}$:
\begin{multline}
\delta_{m}{}^{\prime\prime}+\left[\mathcal{H}-3\mathcal{H}w-\frac{w^{\prime}}{1+w}\right]\delta_{m}^{\prime}\\
+3\left[\mathcal{H}\left(2c_{s}c_{s}^{\prime}-w^{\prime}\right)+\left(c_{s}^{2}-w\right)\left(\mathcal{H}^{\prime}+\mathcal{H}^{2}\left(1-3c_{s}^{2}\right)-\frac{\mathcal{H}w^{\prime}}{\left(1+w\right)}\right)+\frac{k^{2}c_{s}^{2}}{3}\right]\delta_{m}\\
=-\left(1+w\right)\left[3\Phi{}^{\prime\prime}-k^{2}\Phi+3\mathcal{H}\Phi{}^{\prime}\left(1-3c_{s}^{2}\right)\right]\text{ .}\label{eq:rrg-full-eq}
\end{multline}
This equation is clearly too complex to be solved analytically.
In what follows, we make several approximations related to the scales
and periods of interest, imposing matching conditions between solutions
in different eras in order to obtain the solution at late times.

\section{Mészáros equation for WDM\label{Sect3}}

We are interested in studying the sub-horizon evolution of WDM from
the radiation dominated era until now. Before going further with the
perturbation analysis, let us define relevant quantities that will
be used. The equality wave number is given by $k_{\text{eq}}^{2}\equiv\mathcal{H}_{\text{eq}}^{2}$,
so
\begin{equation}
k_{{\rm eq}}\simeq\frac{H_{0}}{a_{{\rm eq}}}\sqrt{2\Omega_{r}}\text{ .}
\label{eq:k-eq}
\end{equation}
As usual, the time variable used in the Mészáros equation will be
\begin{equation}
y=\frac{a}{a_{{\rm eq}}}\,,\label{eq:y-time}
\end{equation}
with $a_{eq}$ given by (\ref{eq:a-equality}).

Considering that the potential is mainly sourced by matter perturbations,
on small scales, Eq.~(\ref{eq:poisson-full}) simplifies to
\begin{equation}
\Phi\simeq\frac{4\pi Ga^{2}}{k^{2}}\rho_{m}\delta_{m}\text{ .}\label{Phi Cap6}
\end{equation}
Using the $y$-variable and, for now, ignoring the contribution of
dark energy, we can express the matter density as
\begin{equation}
4\pi Ga^{2}\rho_{m}\simeq\frac{3\mathcal{H}^{2}}{2}\frac{\sqrt{y^{2}+\frac{b^{2}}{a_{\text{eq}}^{2}}}}{1+\sqrt{y^{2}+\frac{b^{2}}{a_{\text{eq}}^{2}}}}
\ .\label{eq:phi-small-scale}
\end{equation}

Next we assume the warm approximation on the fluid quantities, which
considers the first correction beyond the CDM. Expanding (\ref{eq:EoS}) and (\ref{eq:sound-speed}) for $a\gg b$ and taking the first term
 -- for a detailed consideration of this approximation based on Einstein-Boltzmann equations for RRG see
\cite{Pordeus-da-Silva:2019bak} -- we have
\begin{equation}
w\simeq\frac{b^{2}}{3a_{\text{eq}}^{2}}\frac{1}{y^{2}}\ \text{ and }\ c_{s}^{2}\simeq\frac{5b^{2}}{9a_{\text{eq}}^{2}}\frac{1}{y^{2}}\text{ .}\label{eq:bck-little-warm}
\end{equation}
Moreover, since $a\gg b$ for most of universe evolution, we can simplify
the background evolution according to
\begin{equation}
4\pi Ga^{2}\rho_{m}\simeq\frac{3\mathcal{H}^{2}}{2}\frac{y}{1+y} \ \ \text{and} \ \ \left(\mathcal{H}y\right)^{2}\simeq\frac{H_{0}^{2}\Omega_{r}}{a_{\text{eq}}^{2}}\left(y+1\right)=\frac{k_{\text{eq}}^{2}}{2}\left(y+1\right)\,.\label{eq:hubble-approx}
\end{equation}

Hence, on small scales and considering the warm approximation, the
equation for the evolution of WDM contrast is given by:
\begin{equation}
\partial_{y}^{2}\delta_{m}+\frac{3y+2}{2y\left(y+1\right)}\partial_{y}\delta_{m}-\left(\frac{3}{2}y-\alpha^{2}\right)\frac{\delta_{m}}{y^{2}\left(y+1\right)}=0\text{ },\label{eq:meszaros-warm}
\end{equation}
where 
\begin{equation}
\alpha^{2}=\frac{10}{9}\frac{k^{2}b^{2}}{k_{\text{eq}}^{2}a_{\text{eq}}^{2}}\,.\label{eq:definition-alpha}
\end{equation}
It is important to note that this equation is valid on small scales
($k\gg k_{\text{eq}}$) for values of $y\gg b/a_{\text{eq}}$, as long as dark energy is negligible, later on we will consider the impact of dark energy. The only difference between this equation and the usual Mészáros for CDM \cite{Dodelson,Piattella:2018} is the $\alpha$ term, which encodes the impact of non-negligible sound speed.

It is important to analyze the range of parameters which are valid
when assuming the warm approximation and sub-horizon scales. The horizon
crossing occurs when $k\eta=1$, then 
\begin{equation}
\frac{k}{H_{0}\sqrt{\Omega_{r}}}a_{\text{H}}\simeq 1\text{ \ \ \ }\Rightarrow\text{ \ \ \ }y_{\text{H}}\simeq \frac{H_{0}\sqrt{\Omega_{r}}}{a_{\text{eq}}k}\text{ ,}\label{yH}
\end{equation}
where $y_{\text{H}}\equiv a_{\text{H}}/a_{\text{eq}}$ and $a_{\text{H}}$
is the time a specific mode enters the horizon. We assume that the
equation (\ref{eq:meszaros-warm}) is valid on scales such that
\begin{equation}
y>10y_{\text{H}}\text{ ,}\label{eq:validity-y_h}
\end{equation}
and the warm approximation is valid for 
\begin{equation}
y>20\frac{b}{a_{\text{eq}}}\text{ .}\label{eq:validity-b}
\end{equation}
Combining these two conditions, we get upper limits for $bk$ and $\alpha$:
\begin{equation}
bk<\frac{a_{\text{eq}}k_{\text{eq}}}{2\sqrt{2}}\Rightarrow\alpha<\frac{\sqrt{5}}{6}\simeq0.373.\label{limite}
\end{equation}

\begin{figure}[!htb]
\centering \includegraphics[scale=0.85]{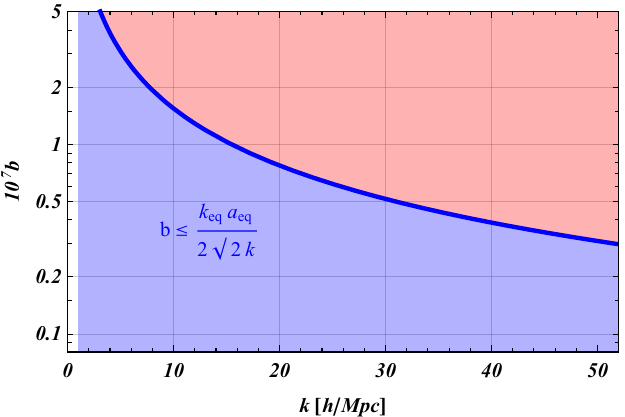} \caption{Allowed parameter space by the warm approximation (blue region), Eq.
(\ref{limite}).}
\label{bfunk}
\end{figure}

In Figure~\ref{bfunk}, we show the allowed range of $b$ as a function
of the wavenumber $k$. The blue area indicates the allowed combination
of $kb$ given by Eq.~(\ref{limite}) and the red region their violation. For instance, considering the lower bound, $m>5.3\ $keV \cite{Irs2017},
which, via Eq.~(\ref{massa b WDM}), corresponds to $b<7.68\times10^{-8}$,
Eq.~(\ref{eq:meszaros-warm}) is valid in the $1\ h \text{/Mpc}<k<20\ h \text{/Mpc}$.

\section{Solving the Mészáros equation\label{Sect4}}

Now that we have discussed the conditions of validity of Eq.~(\ref{eq:meszaros-warm}),
we are going to solve it for epochs of interest and apply the matching
conditions in order to determine the final spectrum of WDM perturbations.
First, we note that Eq.~(\ref{eq:meszaros-warm}) has the following analytical solution:
\begin{equation}
\delta=C_{1}y^{-i\alpha}F\left(-1-i\alpha,\frac{3}{2}-i\alpha,1-2i\alpha;-y\right)+C_{2}y^{i\alpha}F\left(-1+i\alpha,\frac{3}{2}+i\alpha,1+2i\alpha;-y\right),\label{eq:meszaros-an-solution}
\end{equation}
where $F\left(a,b,c;z\right)\equiv$ $_{2}F_{1}\left(a,b,c;z\right)$
is the hypergeometric function, $C_{1}$ and $C_{2}$ are constants.

For $\alpha\neq0$, these solutions are complex, then let us first
determine a real-valued combination of them. The detailed calculation
that determines the real-valued solution is shown in Appendix
\ref{app-A}, the final form is given by:
\begin{equation}
\delta=\bar{C}_{1}\delta_{1_{\operatorname{Re}}}+\bar{C}_{2}\frac{\delta_{2_{\operatorname{Re}}}}{\alpha},\label{eq:meszasros-real-sol}
\end{equation}
where
\begin{multline}
\delta_{1_{\operatorname{Re}}}\equiv2\cos\left(\alpha\ln y\right)\operatorname{Re}F\left(-1-i\alpha,\frac{3}{2}-i\alpha,1-2i\alpha;-y\right)\\
+2\sin\left(\alpha\ln y\right)\operatorname{Im}F\left(-1-i\alpha,\frac{3}{2}-i\alpha,1-2i\alpha;-y\right),\label{delta A1}
\end{multline}
\begin{multline}
\delta_{2_{\operatorname{Re}}}\equiv-2\sin\left(\alpha\ln y\right)\operatorname{Re}F\left(-1-i\alpha,\frac{3}{2}-i\alpha,1-2i\alpha;-y\right)\\
+2\cos\left(\alpha\ln y\right)\operatorname{Im}F\left(-1-i\alpha,\frac{3}{2}-i\alpha,1-2i\alpha;-y\right).\label{delta B1}
\end{multline}
It is straightforward to check that the solution (\ref{eq:meszasros-real-sol})
does not possess the appropriate cold limit, $\lim_{\alpha\rightarrow0}\delta$.
The first part has the expected limit, i.e., 
\[
\lim_{\alpha\rightarrow0}\delta_{1_{\operatorname{Re}}}=2+3y\,,
\]
recovering the usual Mészásros solution for the growing mode. On the other hand, the second part gives 
\[
\lim_{\alpha\rightarrow0}\frac{\delta_{2_{\operatorname{Re}}}}{\alpha}=3\left(2+3y\right)-6\sqrt{1+y}+\left(2+3y\right)\ln\left(\frac{1+\sqrt{1+y}}{2\sqrt{y}}\right)^{2}\ ,
\]
which does not recover the decaying mode of the usual Mészásros solution.

We can circumvent this problem by constructing another solution starting
from the ``good'' one, $\delta_{1}$. Using the Wronskian technique,
the decaying solution which has the expected cold limit is given
by:
\begin{equation}
\delta\left(y\right)=C_{A}\delta_{A}\left(y\right)+C_{B}\delta_{B}\left(y\right)\,,\label{eq:meszaros-final}
\end{equation}
where
\begin{equation}
\begin{split}\delta_{A}\left(y\right)=\frac{2}{3}\cos\left(\alpha\ln y\right)\text{Re}F\left(-1-i\alpha,\frac{3}{2}-i\alpha,1-2i\alpha;-y\right)\\
+\frac{2}{3}\sin\left(\alpha\ln y\right)\text{Im}F\left(-1-i\alpha,\frac{3}{2}-i\alpha,1-2i\alpha;-y\right)
\end{split}
\label{eq:solution-deltaA}
\end{equation}
and 
\begin{equation}
\delta_{B}\left(y\right)=\delta_{A}\left(y\right)\int\frac{dy}{\left[\delta_{A}\left(y\right)\right]^{2}y\sqrt{1+y}} +C\,,
\label{eq:delta-B-integral}
\end{equation}
where the constant $C$ can be determined by demanding that the expression for $\delta _B$ recovers the cold limit in the radiation era -- see Appendix \ref{app-C} for the detailed calculation.
In general, we can not solve (\ref{eq:delta-B-integral}) analytically, but in
the limit $\alpha\rightarrow0$, we get the solution
\begin{equation}
\lim_{\alpha\rightarrow0}\delta_{B}\left(y\right)=\frac{9}{2}\sqrt{1+y}-\frac{9}{4}\left(y+\frac{2}{3}\right)\ln\left(\frac{\sqrt{1+y}+1}{\sqrt{1+y}-1}\right)\,,\label{eq:solution-deltaB}
\end{equation}
which recovers the decaying solution for CDM for $y\gg1$
\begin{equation}
\lim_{\alpha\rightarrow0}\delta_{B}\left(y\right)\propto y^{-3/2}\,.\label{eq:delta-B-CDM-late}
\end{equation}
Moreover, we have verified numerically that the solution (\ref{eq:delta-B-integral})
is decaying for $\alpha>0$. Hence, the solution (\ref{eq:meszaros-final})
is the relevant one for the WDM Mészáros equation.

\subsection{Initial conditions and horizon crossing}

Initially the universe is dominated by radiation, then the potential satisfies
\begin{equation}
\Phi^{\prime\prime}+\frac{4}{\eta}\Phi^{\prime}+\frac{k^{2}}{3}\Phi=0\,.\label{eq:phi-2-radiation}
\end{equation}
On large-scales, the non-decaying solution is constant $\Phi=\Phi_{{\rm p}}$.
As usual, we set adiabatic initial conditions on large-scales, related
to the primordial value of the potential
\begin{equation}
\frac{\delta_{m,\text{p}}}{3\left(1+w_{\text{p}}\right)}=\frac{\delta_{r,\text{p}}}{4}=\frac{1}{2}\Phi_{\text{p}}.
\end{equation}
As the universe expands, the large-scale modes enter the horizon and
the general solution of Eq.~(\ref{eq:phi-2-radiation}) is given by
\begin{equation}
\Phi(k,\eta)=3\Phi_{\text{p}}\left[\frac{\sin(k\eta/\sqrt{3})-(k\eta/\sqrt{3})\cos(k\eta/\sqrt{3})}{(k\eta/\sqrt{3})^{3}}\right]\,.\label{eq: phi-rad}
\end{equation}
For CDM, the solution of Eq.~(\ref{eq:rrg-full-eq}), with the potential
given by (\ref{eq: phi-rad}), for modes inside the horizon is \cite{Dodelson,Piattella:2018}
\begin{equation}
\delta_{c}\left(x\right)=A\Phi_{\text{p}}\ln\left(Bx\right)\,,\label{eq:delta-cold-small}
\end{equation}
where $A=9$, $B\simeq0.6237$ and $x=k\eta$ (or  $x=y/y_{H}$ during the radiation dominated era).

In order to have a clear correspondence between CDM and WDM in the
cold limit, based on the form of the analytical solution of Mészáros equation,
Eq.~(\ref{eq:meszaros-final}), we propose the following ansatz for
the warm case
\begin{align}
\delta_{m}\left(x\right) & =\frac{A\Phi_{\text{p}}}{\alpha}\sin\left[\alpha\ln\left(Bx\right)\right]\text{ , \ \ }(x\gg1)\,.
\label{eq: warm-ansatz}
\end{align}
More details about the motivation for choosing this
expression are given in Appendix \ref{app-B}. In practice, expanding (\ref{eq: warm-ansatz})
for small $\alpha\ln\left(Bx\right)$ is accurate and will be useful
to determine matching conditions between epochs,
\begin{equation}
\delta_{m}\left(x\right)\simeq A\Phi_{\text{p}}\ln\left(Bx\right)\left[1-\frac{1}{6}\alpha^{2}\ln^{2}\left(Bx\right)\right]\text{ , \ \ }(x\gg1\text{ and }\alpha\ln\left(Bx\right)\ll1)\,.\label{eq: warm-ansatz-expansion}
\end{equation}

\begin{figure}[!htb]
\includegraphics[width=1\columnwidth]{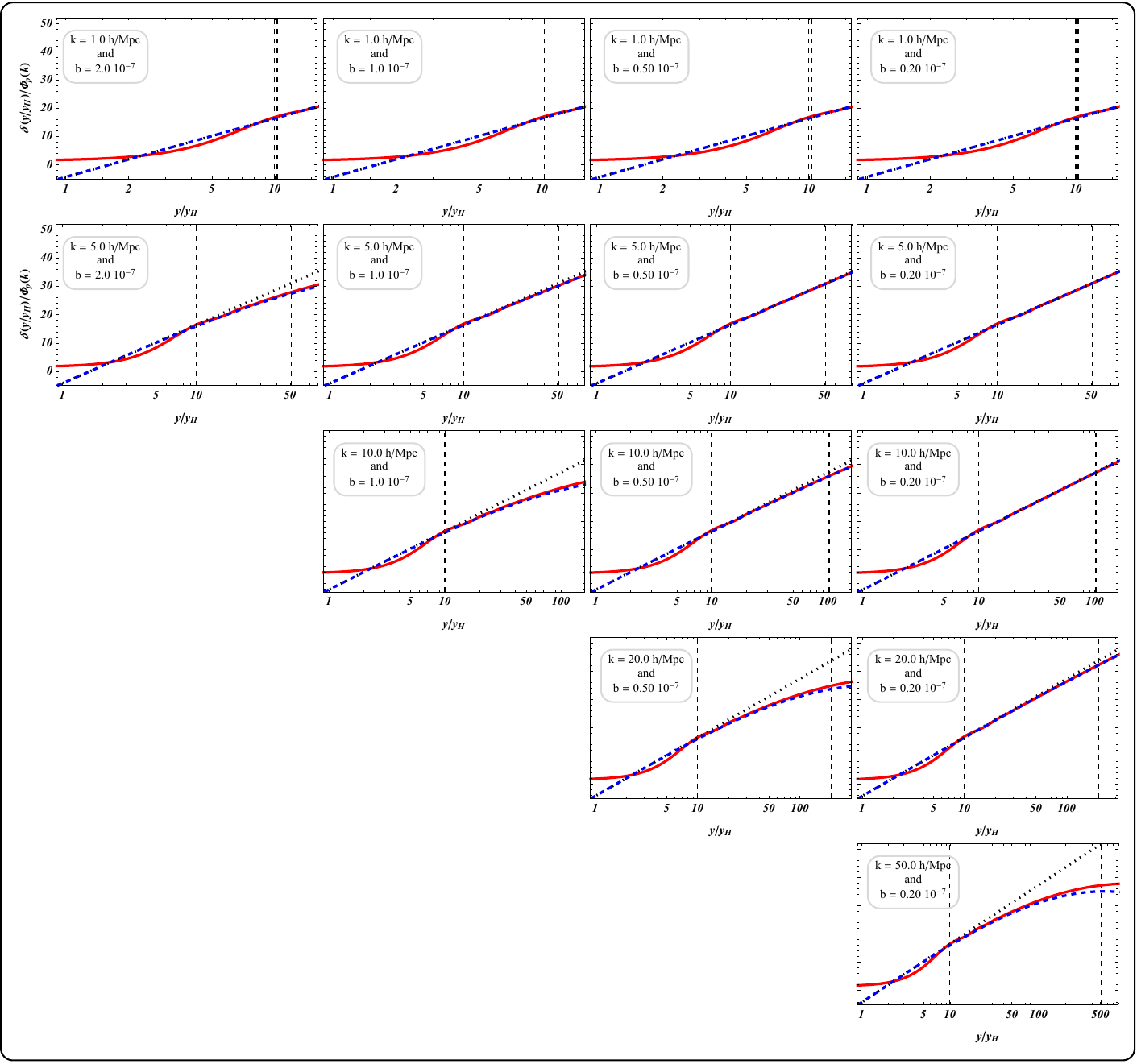}
\caption{Evolution of WDM perturbations in the radiation dominated era according to the numerical solution of Eq.~(\ref{Eq Dif delta rrg}) (red line) and the ansatz (\ref{eq: warm-ansatz-expansion}) (blue dashed line).
The black dotted line indicates the solution for CDM, (\ref{eq:delta-cold-small}). The vertical lines indicate the range $10y_{\text{H}}\lesssim y\lesssim10^{-1}$, in which the analytical solutions are valid. 
We have only used values of $k$ and $b$ allowed by Eq.~(\ref{limite}),
shown in Figure~\ref{bfunk}.}
\label{fig:ansatz-numerical}
\end{figure}

Now let us compare (\ref{eq: warm-ansatz-expansion}) with the corresponding
numerical solution. Deep in the radiation era, Eq.~(\ref{eq:rrg-full-eq})
can be written as
\begin{equation}
k^{2}\partial_{x}^{2}\delta_{m}{}+kF(k,x)\partial_{x}\delta_{m}{}+G(k,x)\delta_{m}{}=S(k,x)\text{ ,}
\label{Eq Dif delta rrg}
\end{equation}
where $G(k,x)$, $F(k,x)$ and $S(k,x)$, are given by:
\begin{equation}
G(k,x)\equiv3\left(\frac{k}{x}\right)^{2}\left[2c_{s}x\partial_{x}c_{s}-x\partial_{x}w-\left(c_{s}^{2}-w\right)\left(3c_{s}^{2}+\frac{x\partial_{x}w}{\left(1+w\right)}\right)+\frac{x^{2}c_{s}^{2}}{3}\right]\ ,
\end{equation}
\begin{equation}
F(k,x)\equiv\frac{k}{x}\left[1-3w-\frac{x\partial_{x}w}{1+w}\right]\ ,
\end{equation}
\begin{equation}
S(k,x)=k^{2}\left(1+w\right)\left[9\frac{\left(1+c_{s}^{2}\right)}{x}\partial_{x}\Phi{}{}+2\Phi\right]\text{ .}
\end{equation}
We solve Eq.~(\ref{Eq Dif delta rrg}) numerically and compare the
results with the ansatz (\ref{eq: warm-ansatz-expansion}). One can
see in Figure~\ref{fig:ansatz-numerical} that they are in very good
accordance. The interval between the vertical lines in this figure
represents the range $10a_{\text{H}}\lesssim a\lesssim10^{-1}a_{\text{eq}}$,
which corresponds to modes well inside the horizon and deep in the
radiation era. As we can see, the numerical solution (red line) and
the ansatz (blue-dashed line) are in good accordance slightly after
the horizon crossing and even beyond $a=10^{-1}a_{\text{eq}}$.

In Figure~\ref{fig:ansatz-numerical} we can see the most important
effect of warmness. For a given scale, increasing $b$ damps the growth
of dark matter perturbations. Conversely, for a given $b$, the warmness
is more important for smaller scales. Hence, during the radiation
dominated era, the warmness induces a scale-dependent decrease in
dark matter growth.

In order to determine solutions after the horizon crossing, we have to set the constants $C_{A}$ and $C_{B}$ in solution (\ref{eq:meszaros-final})
by demanding it is compatible with the ansatz (\ref{eq: warm-ansatz-expansion}).
Details of this procedure are given in Appendix \ref{app-B}, the results are:
\begin{equation}
C_{A}=\frac{3}{2}A\Phi_{\text{p}}\left\{ \ln\left(4e^{-3}\frac{B}{y_{\text{H}}}\right)-\frac{\alpha^{2}}{6}\ln^{2}\left(\frac{B}{y_{\text{H}}}\right)\left[\ln\left(\frac{B}{y_{\text{H}}}\right)-3\ln\left(\frac{e^{3}}{4}\right)\right]\right\} \label{eq:CA-sol}
\end{equation}
and 
\begin{equation}
C_{B}=\frac{2}{3}A\Phi_{\text{p}}\left\{ 1-\frac{\alpha^{2}}{2}\ln^{2}\left(\frac{B}{y_{\text{H}}}\right)\right\} \,.\label{eq:CB-sol}
\end{equation}
Note that the constants depend on $\alpha$, which encodes the importance
of warmness on a given scale. The solution $C_{B}\delta_{B}\left(y\right)$
is decaying and $C_{A}\delta_{A}\left(y\right)$ is the growing solution.
Thus the WDM solution is given by (\ref{eq:meszaros-final}), where
the constants are given by (\ref{eq:CA-sol}) and (\ref{eq:CB-sol}).

\subsection{Solution in the matter era}

\begin{figure}[!htb]
\centering \includegraphics[scale=1.2]{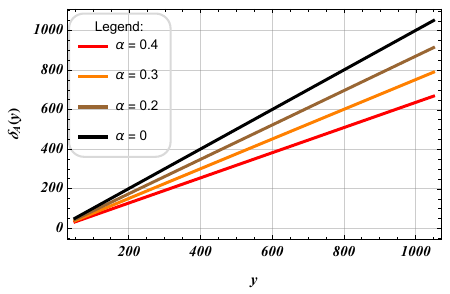} \caption{Evolution of analytical solution $\delta_{A}$ in
the matter era, Eq.~(\ref{eq:solution-deltaA}),
for different warmness parameters $\alpha$. The case with $\alpha=0$
corresponds to the usual CDM solution.}
\label{FigdA}
\end{figure}

Now let us check the behavior of solution (\ref{eq:meszaros-final})
deep in the matter dominated era, $y\gg1$. The $\delta_{B}$ part
is decaying, then the relevant solution is
\begin{equation}
\delta_{m}(y,\alpha)\simeq C_{A}\delta_A(y,\alpha)\,,\label{eq:solution-rad-mat}
\end{equation}
where $C_{A}$ and $\delta_{A}$ are given by (\ref{eq:CA-sol}) and (\ref{eq:solution-deltaA}), respectively. As can be seen in Figure~\ref{FigdA}, the function $\delta_{A}$ is essentially a linear
function whose slope depends on $\alpha$. Hence,
the growth is essentially the same as for that the cold dark matter case, $\delta_{m}\propto a$. This happens because, for typical warmness
parameters, the fluid is already very cold in the matter dominated
era. However, the impact of warmness is present in the initial values
of perturbations on the onset of matter dominance, which depends on
the free streaming scale at earlier times in the radiation dominated era.
Therefore the main impact of warmness occurs in the early universe,
and the evolution after the matter-radiation equality is essentially
the same as that of CDM.

Clearly, we can simplify the solution (\ref{eq:solution-deltaA})
by fitting a linear function with coefficients that depend on $\alpha$:
\begin{equation}
\tilde{\delta}_{A}\left(\alpha,y\right)\simeq yf\left(\alpha\right)+g\left(\alpha\right)\,.\label{eq:ansatz-lin}
\end{equation}
Demanding that the Mészáros solution is recovered in the cold limit,
we must have:
\begin{equation}
\lim_{\alpha\rightarrow0}f\left(\alpha\right)=1\text{ \ and \ \ }\lim_{\alpha\rightarrow0}g\left(\alpha\right)=\frac{2}{3}\text{.}\label{eq:limits-line-coeff}
\end{equation}
By numerical exploration, we found the following fit to the coefficients:
\begin{equation}
f\left(\alpha\right)=\frac{1}{1+3.806\alpha^{2}-1.418\alpha^{4}}\label{eq:coef-f-alpha}
\end{equation}
and 
\begin{equation}
g\left(\alpha\right)=\frac{2/3}{1+2.783\alpha^{2}-3.603\alpha^{4}}.\label{eq:coef-g-alpha}
\end{equation}
We have checked that the determination of these coefficients is independent
of $y$ for $y>100$. These expressions reproduce the solution (\ref{eq:solution-deltaA}) with precision better than $0.02\%$.

Finally, the evolution of WDM perturbations deep in the matter era is
well described by
\begin{equation}
\delta_{m}\left(y,k,\alpha\right)\simeq C_{A}\left(k,\alpha\right)\tilde{\delta}_{A}\left(y,\alpha\right)\label{eq:solution-deep-matter-scale-dep}
\end{equation}
where $C_{A}$ is given by Eq.~(\ref{eq:CA-sol}) and $\tilde{\delta}_{A}$
is an approximate representation of (\ref{eq:solution-deltaA}) for
$y\gg1$.

\section{Impact of other components\label{Sect5}}

\subsection{Dark Energy}

Let us now consider the impact of dark energy in the form of the Cosmological Constant, $\Lambda$. As just described, the growth of WDM perturbations in the matter dominated era is already effectively scale invariant, therefore the following analysis is the same as that for CDM in the dark energy dominated era, \cite{Dodelson,ellis_maartens_maccallum_2012}. The background evolution at late times is now given by
\begin{equation}
H^{2}\simeq H_{0}^{2}\left(\Omega_{\Lambda}+\Omega_{m}a^{-3}\right)\label{eq:hubble-dark-energy}
\end{equation}
Note that when $\Lambda$ becomes important, the evolution of matter
is essentially cold, see Figure~\ref{fig:background}. At these late times,
we can also consider that matter is cold at the perturbation level,
then we get the usual equation for matter growth on small scales
\begin{equation}
\frac{d^{2}\delta_{m}}{da^{2}}+\left(\frac{3}{a}+\frac{1}{H}\frac{dH}{da}\right)\frac{d\delta_{m}}{da}-\frac{3H_{0}^{2}\Omega_{m}}{2a^{5}H^{2}}\delta_{m}=0.\label{eq:matter-growth-lambda}
\end{equation}

One can check that $\delta_{m}\propto H$ is a solution of Eq.~(\ref{eq:matter-growth-lambda}).
Then we construct the second solution using the Wronskian. The complete solution is given by
\begin{equation}
\delta_{m}\left(a\right)\simeq C_{1}H\int\frac{1}{\left(aH\right)^{3}}da+C_{2}H\left(a\right)\,.\label{eq:solution-matter-Lambda}
\end{equation}
Now we need to set the new constants, $C_{1}$ and $C_{2}$. Deep
in the matter era, (\ref{eq:matter-growth-lambda}) has the well known
solution
\begin{equation}
\delta_{m}\simeq C_{1}a+C_{2}a^{-3/2}\,.\label{eq:solution-deep-matter-scale-ind}
\end{equation}
By demanding that this expression is compatible with (\ref{eq:solution-deep-matter-scale-dep}) and neglecting the $g(\alpha)$ term,
we get
\[
C_{1}=\frac{5H_{0}^{2}\Omega_{m}}{2}\frac{C_{A}\left(k,\alpha\right)f\left(\alpha\right)}{a_{eq}}\text{ and }C_{2}=0.
\]

Finally, solving the integral in (\ref{eq:solution-matter-Lambda})
we get the final solution for WDM at low-$z$
\begin{equation}
\delta_{m}\left(a\right)=\frac{5C_{A}\left(k,\alpha\right)f\left(\alpha\right)}{3a_{eq}}a\left[1-\frac{2}{5}\left(a^{3}\frac{\Omega_{\Lambda}}{\Omega_{m}}+1\right)^{1/2}F\left(\frac{1}{2},\frac{5}{6},\frac{11}{6};-a^{3}\frac{\Omega_{\Lambda}}{\Omega_{m}}\right)\right].\label{eq:delta-m-final}
\end{equation}
A simpler form can be obtained by expanding the solution for $a<1$,
\begin{equation}
\delta_{m}\left(a\right)\simeq C_{A}\left(k,\alpha\right)f\left(\alpha\right)\frac{a}{a_{eq}}\left[1-\frac{2}{11}\frac{a^{3}\Omega_{\Lambda}}{\Omega_{m}}+\frac{16}{187}\left(\frac{a^{3}\Omega_{\Lambda}}{\Omega_{m}}\right)^{2}+\mathcal{O}\left(\left(\frac{a^{3}\Omega_{\Lambda}}{\Omega_{m}}\right)^{3}\right)\right]\,,\label{eq:final-solution-approx}
\end{equation}
which, at $a=2/3$ ($z=1/2$), differs from (\ref{eq:delta-m-final})
by about $4\%$.

\subsection{Baryons}

So far we have considered that matter is composed by WDM and baryons
without distinction. However, we know that baryons are coupled to photons
until around $z\simeq1100$ and their clustering is possible only afterwards.
We can correct this effect with the approach presented in \cite{Hu:1995en}.
The matter perturbation is then rescaled according to
\begin{equation}
\delta_{m}=\beta\left(1-\frac{\Omega_{b}}{\Omega_{m}}\right)\times\lim_{\frac{\Omega_{b}}{\Omega_{m}}\rightarrow0}\delta_{m}\ ,\label{add_baryons}
\end{equation}
where $\lim_{\frac{\Omega_{b}}{\Omega_{m}}\rightarrow0}\delta_{m}$
is given by (\ref{eq:delta-m-final}), $\Omega_{m}=\Omega_{dm}+\Omega_{b}$
and $\beta=(47\Omega_{m}h^{2})^{-0.67\Omega_{b}/\Omega_{m}}$, which
is valid for $\Omega_{b}/\Omega_{m}\leq0.5$. Although this correction
was first proposed in the context of CDM, it should be valid for WDM
as well, because the main effect encoded in Eq.~(\ref{add_baryons})
comes from photon-baryon interactions, which is unaffected by the
nature of dark matter.

\subsection{Final form of the solution}

Putting together all the analysis for WDM perturbations, the complete
solution valid for an era of matter domainance and dark energy is given by
\begin{equation}
\delta_{m}\left(b,k,z\right)=\beta\left(1-\frac{\Omega_{b}}{\Omega_{m}}\right)\times C_{A}\left(k,\alpha\right)\times f\left(\alpha\right)\frac{1+z_{\text{eq}}}{1+z}\times g_{\Lambda}(z)\ ,\label{delta_total_final}
\end{equation}
where
\begin{align}
f\left(\alpha\right) & \simeq\frac{1}{1+3.806\alpha^{2}-1.418\alpha^{4}}\,,\label{falpha}\\
C_{A}\left(k,\alpha\right) & =\frac{3}{2}A\Phi_{\text{p}}\left\{ \ln\left(4e^{-3}\frac{B}{y_{\text{H}}}\right)-\frac{\alpha^{2}}{6}\ln^{2}\left(\frac{B}{y_{\text{H}}}\right)\ln\left(4^{3}e^{-9}\frac{B}{y_{\text{H}}}\right)\right\} \,,\\
g_{\Lambda}\left(z\right) & =\frac{5}{3}-\frac{2}{3}\left(\frac{\Omega_{\Lambda}}{\Omega_{m}\left(1+z\right)^{3}}+1\right)^{1/2}F\left(\frac{1}{2},\frac{5}{6},\frac{11}{6};-\frac{\Omega_{\Lambda}}{\Omega_{m}\left(1+z\right)^{3}}\right)\,.
\end{align}
In order to perform a simple consistency test, we calculated the relative deviation between our function in its cold limit ($b=0$) and it's equivalent obtained via numerical calculation through CAMB, assuming the standard $\Lambda$CDM scenario. From this test, we conclude that the relative deviation is around $1\%$ for $k>4 \ h\text{/Mpc}$ and slightly increases at larger scales, reaching $4\%$ at $k=1 \ h\text{/Mpc}$. Furthermore, it is observed that this result depends very little on the value of $z$, as long as it is in the range of allowed values, that is, $z\ll z_\text{eq}$.

\section{Power spectrum and transfer function\label{Sect6}}

With the solution (\ref{delta_total_final}), we can compute the matter power spectrum
\begin{equation}
P_{m}\left(b,k,z\right)=|\delta_{m}\left(b,k,z\right)|^{2}\,,
\end{equation}
on small scales. Neglecting neutrino multipoles higher than $\ell=2$,
the curvature perturbation $\mathcal{R}$ is related to the Newtonian
potential by $\Phi_{\text{p}}(k)=\frac{2}{3}\mathcal{R}$. Then we
assume that the primordial power spectrum of $\Phi$ is given
by
\begin{equation}
\Phi_{\text{p}}^{2}(k)=\frac{8}{9}\pi^{2}A_{s}k^{-3}\left(\frac{k}{k_{*}}\right)^{n_{s}-1}\,,
\end{equation}
where $A_{S}$ is the amplitude of scalar perturbations, $n_{s}$
the primordial spectral index, $k_{*}=0.05/\text{Mpc}$ the pivot
scale used in Planck analysis \cite{Aghanim:2018eyx}.

\begin{figure}[!htb]
\centering{}\includegraphics[scale=0.85]{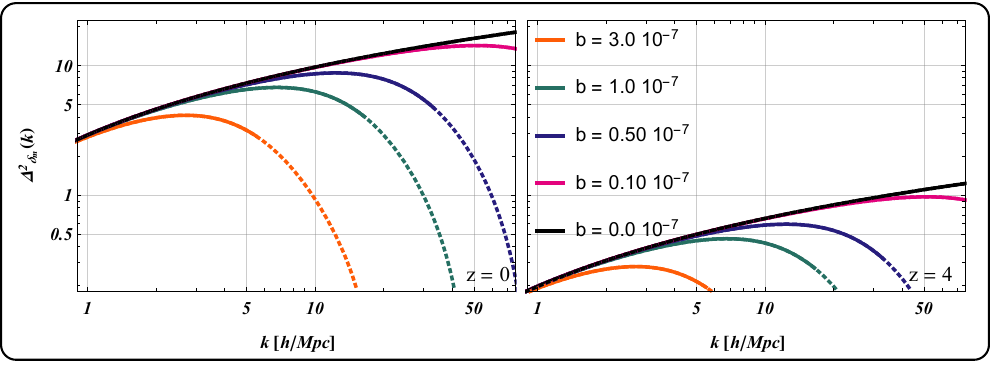}\caption{Dimensionless matter power spectrum, $\Delta_{m}^{2}$, as a function
of $k$ for two redshifts and several values of $b$ indicated in
the legend of the right plot. The continuous lines become dotted when
the warm approximation breaks. Note that, besides the violation of
this approximation, the behaviour of $\Delta_{m}^{2}$ is consistent
with the expected damping of WDM perturbations on small scales. The condition (\ref{eq:validity-b}) is arbitrary and could
be alleviated, allowing the solutions to be continued to slightly smaller scales without introducing large errors. \label{fig:PS-adimensional}}
\end{figure}

The dimensionless matter power spectrum is given by
\begin{equation}
\Delta_{m}^{2}\left(b,k,z\right)=\frac{k^{3}P_{m}\left(b,k,z\right)}{2\pi^{2}}\,.
\end{equation}
When $\Delta_{m}^{2}>1$, the linear evolution of perturbations breaks
down, which also indicates the time and scales of nonlinear structure
formation. In Figure~\ref{fig:PS-adimensional}, we show the dependence of
the dimensionless matter power spectrum on the warmness parameter
$b$ for three different redshifts.

\begin{figure}[!htb]
\centering{}\includegraphics[scale=0.6]{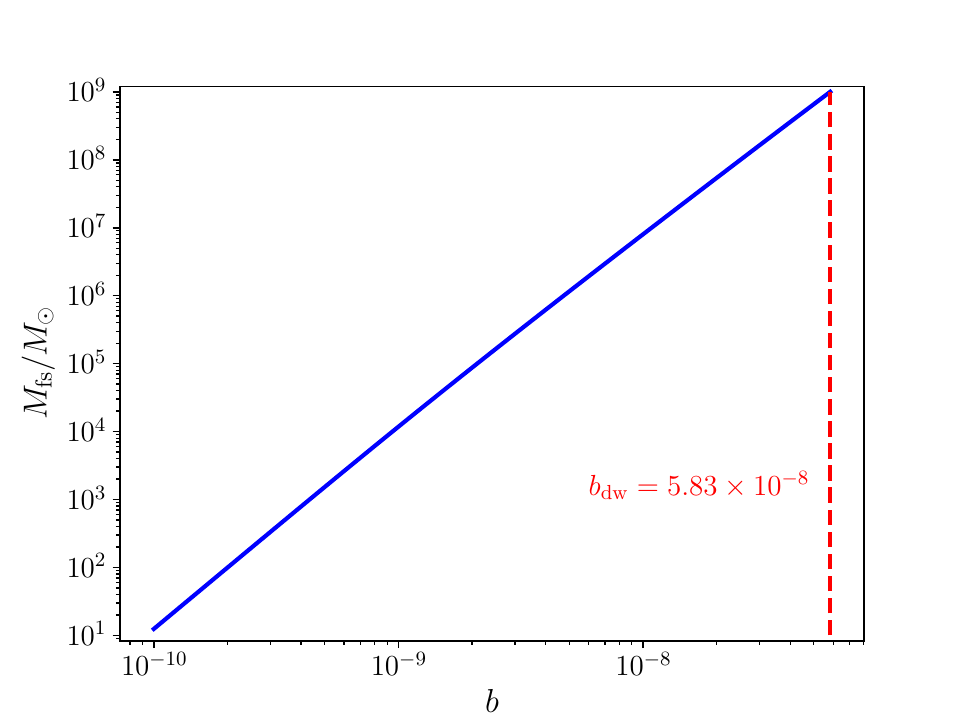}\caption{Free streaming mass scale as a function of $b$ (blue line) and $b$
value at the scale of dwarf galaxies mass, $M=10^{9}M_{\odot}$ (red
dashed line).\label{fig:free-stream-mass}}
\end{figure}

As can be seen, increasing $b$ damps the growth of perturbations
on small scales, and it is clear that WDM has the potential to suppress the formation of small halos, possibly alleviating the Missing Satellites Problem, e.g.,  Ref.~\cite{Angulo2013,Lovell2014,Horiuchi2016}.
Another related quantity that can be easily and accurately calculated
with the RRG model for WDM is the free-streaming mass scale:
\begin{equation}
M_{{\rm fs}}=\frac{4\pi}{3}r_{{\rm fs}}^{3}\rho_{m}\,,\label{eq:free-stream-mass}
\end{equation}
where
\begin{equation}
r_{{\rm fs}}=\int_{a_{i}}^{1}\frac{c_{s}\left(a\right)da}{a^{2}H}\,.\label{eq:free-stream-radius}
\end{equation}
In Figure~\ref{fig:free-stream-mass} we show the dependence of $M_{{\rm fs}}$
on $b$. The value of $b$ corresponding to the mass scale of dwarf galaxies,
$10^{9}M_{\odot}$, $b_{{\rm dw}}=5.83\times10^{-8}$ is also shown.
The actual abundance of small dark matter halos is indeed strongly
dependent on the free steaming mass scale Ref.~\cite{Angulo2013}. Thus, assuming that halos below some mass are not formed,
this simple calculation gives an upper limit for $b$. Note that this estimate is independent of the approximations used to determine the evolution of WDM perturbations.

\begin{figure}
\centering{}\includegraphics[scale=1.2]{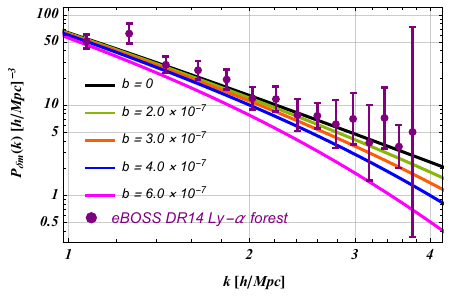}\caption{Matter power spectrum for several values of $b$ and matter power spectrum data derived from Lyman-$\alpha$
forest observations from Ref.~\cite{Chabanier:2019eai}. \label{fig:PS-Lyman-a}}
\end{figure}

In Figure~\ref{fig:PS-Lyman-a}, we also show the matter power spectrum
at $z=0$ and the data derived from Lyman-$\alpha$ forest observations
\cite{Chabanier:2019eai}. Although this data is obtained assuming
the $\Lambda$CDM model, we use it to test WDM in the context of a
consistency check. As can be seen in this example, increasing the
warmness beyond $b=4\times10^{-7}$ is incompatible with the data
shown in the plot.

\begin{figure}[!htb]
\centering{}
\includegraphics[scale=0.75]{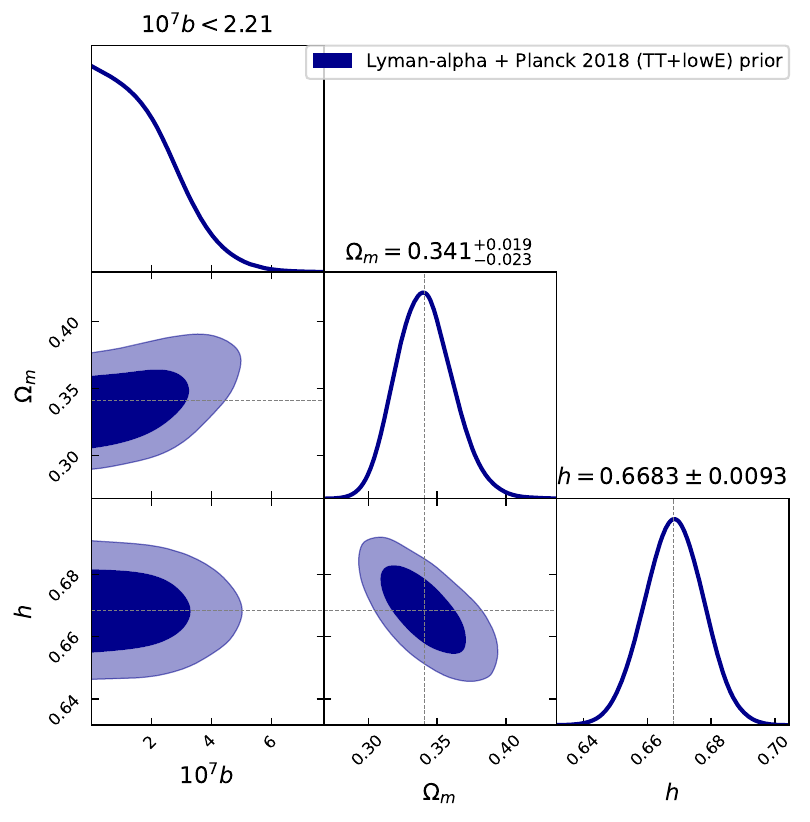}
\caption{Constraints on the warmness parameter $b$ using Gaussian priors $A_{S}$, $n_{s}$, $\Omega_{b}h^{2}$ and $h$ derived from Planck 2018 results, leaving $\Omega_{m}$ and $b$ free (flat
uninformative priors). We stress again that this is only an exercise to verify the consistency of our solutions and explore the possible dependence
of $b$ with the cosmological model. 
\label{fig:triangle-plot}}
\end{figure}

As another exercise to test the consistency of our result, we evaluate
the constraints on the parameter $b$ allowed by data of \cite{Chabanier:2019eai}.
We stress that the purpose of this analysis is to test the consistency
of our solution and not to provide realistic statistical constraints
on the warmness of DM. For the parameters $A_{S}$, $n_{s}$, $\Omega_{b}h^{2}$ and $h$ we assume Gaussian priors based on Planck 2018 results \cite{Aghanim:2018eyx}. The results for a selected combination
of parameters are shown in Figure~\ref{fig:triangle-plot}. To obtain these results, we made use of the Monte Carlo technique and statistical analysis using the respective public packages \href{https://emcee.readthedocs.io/en/stable/}{emcmc} \cite{emcee} and \href{https://getdist.readthedocs.io/en/latest/intro.html}{GetDist} \cite{Lewis:2019xzd}.

\begin{figure}[!htb]
\centering{}\includegraphics[scale=1.2]{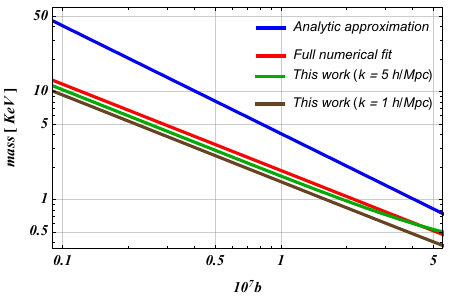}\caption{Relation between the particle's mass and the warmness parameter $b$:
analytic modeling (blue line) \cite{Pordeus-da-Silva:2019bak}, fit
from numerical transfer function \cite{RRG2018Hipolito} (red line)
and fits using the analytic transfer function found in this work (green
and brown lines), Eq.~(\ref{eq:rrg-mass}). As can be seen, the analytic
and numerical fits based on the transfer function are in good agreement.
\label{fig:b-mass}}
\end{figure}

With the analytic solution for WDM perturbations, we can also evaluate
the relative transfer function, defined by
\begin{equation}
T\left(k\right)=\left(\frac{P_{WDM}}{P_{CDM}}\right)^{1/2}\,.
\end{equation}
A popular fit is given by \cite{Bode_2001}
\begin{equation}
T\left(k\right)=\left[1+\left(\mu k\right)^{2}\right]^{-5/\nu}\,,\label{eq:transfer-bode}
\end{equation}
where 
\[
\mu=0.049\left(\frac{m}{1\ \text{kev}}\right)^{-1.11}\left(\frac{\Omega_{m}}{0.25}\right)^{0.11}\left(\frac{h}{0.7}\right)^{1.22}h^{-1}\text{Mpc}
\]
and $\nu=1.12$ for $k<5\ h \text{/Mpc}$ \cite{Viel:2005qj}.
We compute our transfer function by setting $P_{CDM}=P_{m}\left(b=0\right)$
and $P_{WDM}=P_{m}\left(b\right)$, which gives
\begin{equation}
T_{RRG}=\frac{C_{A}\left(k,\alpha\right)f\left(\alpha\right)}{C_{A}\left(k,0\right)}\label{eq:transfer-rrg}
\end{equation}
Then, comparing it to the previous expressions we can determine an
expression for the mass of RRG particles 
\begin{equation}
m_{RRG}=\frac{\left[0.049\left(\frac{k\ \text{Mpc}}{h}\right)\left(\frac{\Omega_{m}}{0.25}\right)^{0.11}\left(\frac{h}{0.7}\right)^{1.22}\right]^{1/1.11}}{\left[\left(\frac{C_{A}\left(k,0\right)}{C_{A}\left(k,\alpha\right)f\left(\alpha\right)}\right)^{\nu/5}-1\right]^{1/(2.22\nu)}}\ \text{keV}\,.\label{eq:rrg-mass}
\end{equation}
Note that the mass depends on cosmological parameters and also on
$k$. This is a formal inconsistency (the particle's mass should be
unique). However, we must note that (\ref{eq:transfer-bode}) is a numerical
fit, which can not be totally accurate, and our solution is derived
under some approximations. Nevertheless, the dependence of (\ref{eq:rrg-mass})
on $k$ is very small, and the $b$-mass relation is very close to
the one found in \cite{RRG2018Hipolito} using a numerical fit, see
Figure~\ref{fig:b-mass}.

Regarding the $b$-mass relation, one should bear in mind that the
main parameter in the RRG is $b$, which is directly related to the
thermal velocities of particles, which, in turn, is the most relevant quantity that dictates the evolution of perturbations. The association of $b$
with the particle mass is model dependent, as discussed in \cite{Pordeus-da-Silva:2019bak}.
As we see from (\ref{eq:rrg-mass}), the numerical fit to the $b$-mass
relation also depends on some assumptions. Therefore, when using the
RRG model, the fundamental quantity to report is $b$ or, via Eq.~(\ref{eq:EoS}),  the
thermal velocity, $v_{th}$. Using the result shown in Figure~\ref{fig:triangle-plot}, our
simple constraint exercise, we find
$v_{th}<66.3\ \text{m}/\text{s}$ now. This value is very close to the one
reported in Ref.~\cite{Armendariz-Picon:2013jej}, which used an approximated
Maxwell distribution, $v_{th}<54\ \text{m}/\text{s}$. In terms of mass, using the expression $m=4.65\times10^{-6}b^{-4/5}\ \text{keV}$ from Ref.~\cite{RRG2018Hipolito}, we get $m>0.981\ \text{keV}$.

\section{Conclusions\label{Sect7}}

In this work, we have used the RRG model to describe the WDM perturbations
on small scales under the warm approximation of RRG. We derived a
Mészáros-like equation which takes into account the effect of non-null
velocity dispersion of dark matter particles and analytically solved
it for radiation, matter and dark energy dominated eras, matching
the different solutions and also considering the distinct behaviour
of baryons.

Our analysis shows that, deep in the radiation era, the growth of matter perturbations is suppressed by the non-null velocity dispersion of the constituent particles in a scale-dependent way, encoded by $\alpha$. When the WDM perturbations enter the matter dominated era, they have already cooled and then grow linearly, just like in the CDM case, but starting at values that depend on $\alpha$. Hence, the growth of WDM becomes scale invariant after matter-radiation equality, because the remaining effects of warmness are negligible on scales that we have considered, $1\ h \text{/Mpc}\lesssim k\lesssim10\ h \text{/Mpc}$.

With our analytical solution for WDM perturbations, we have constructed the matter power spectrum and the relative transfer function on small
scales. We have also performed a simplified statistical analysis to constrain
the warmness parameter, $b$, giving $b<2.21\times10^{-7}$ (2$\sigma$
C.L.) or, equivalently for the thermal velocity now, $v_{th}<66.3\text{m}/\text{s}$, in good accordance with Ref.~\cite{Armendariz-Picon:2013jej}.

Our developments show that the simplicity of the RRG model offers a valuable
opportunity to analytically solve and understand the evolution of
WDM linear perturbations throughout the whole cosmic history, while
it is still capable of providing meaningful estimates for the degree
of the warmness of dark matter particles.

\acknowledgments
LGM acknowledges CNPq-Brazil (grant 308380/2019-3) for partial financial support.

\appendix
\section{Real solutions}\label{app-A}

Here we give some details about how to obtain the real solution of
Eq.~(\ref{eq:meszaros-warm}), shown in Eq.~(\ref{eq:meszasros-real-sol}),
i.e.,
\begin{equation}
\delta\left(y\right)=C_{1}\delta_{1}\left(y\right)+C_{2}\delta_{2}\left(y\right)\,,
\end{equation}
where
\begin{equation}
\delta_{1}\left(y\right)\equiv y^{-i\alpha}F\left[-1-i\alpha,\frac{3}{2}-i\alpha,1-2i\alpha;-y\right]\text{ \ \ and \ \ }\delta_{2}\left(y\right)\equiv y^{i\alpha}F\left[-1+i\alpha,\frac{3}{2}+i\alpha,1+2i\alpha;-y\right].\label{delta 1-1}
\end{equation}
Using 
\begin{align}
y^{ia} & =\cos\left(\alpha\ln y\right)+\text{\ }i\sin\left(\alpha\ln y\right),\label{yplus}\\
y^{-ia} & =\cos\left(\alpha\ln y\right)-\text{\ }i\sin\left(\alpha\ln y\right),\label{yminus}
\end{align}
and making use of the following relations: 
\begin{align}
\operatorname{Re}\left\{ F\left[-1-i\alpha,\frac{3}{2}-i\alpha,1-2i\alpha;-y\right]\right\}  & =\operatorname{Re}\left\{ F\left[-1+i\alpha,\frac{3}{2}+i\alpha,1+2i\alpha;-y\right]\right\} ,\label{Real}\\
\operatorname{Im}\left\{ F\left[-1-i\alpha,\frac{3}{2}-i\alpha,1-2i\alpha;-y\right]\right\}  & =-\operatorname{Im}\left\{ F\left[-1+i\alpha,\frac{3}{2}+i\alpha,1+2i\alpha;-y\right]\right\} .\label{Imaginario}
\end{align}
Then two real valued solutions can be obtained choosing $C_{1}\rightarrow C_{2}=1$
e $C_{2}\rightarrow-C_{1}=i$.

\section{Ansatz choice}\label{app-B}

In this appendix, we explain the considerations which motivate the
ansatz (\ref{eq: warm-ansatz}). Deep in the radiation era, $y\ll1$,
the leading term in Eq.~(\ref{eq:solution-deltaA})
\begin{equation}
\delta_{A}\approx\frac{2}{3}\cos\left(\alpha\ln y\right).
\end{equation}
Substituting this result in Eq.~(\ref{eq:delta-B-integral}), we get
\begin{equation}
\delta_{B}\approx\frac{3}{2}\frac{\sin\left(\alpha\ln y\right)}{\alpha}\text{ ,\,\,\,}y\ll1.
\end{equation}
Hence, deep in the radiation era, we have the simplified solution
\begin{equation}
\delta_{m}\left(y\right)\approx\bar{C}_{A}\cos\left(\alpha\ln y\right)+\bar{C}_{B}\frac{\sin\left(\alpha\ln y\right)}{\alpha}.\label{eq:appen-delta-m-rad}
\end{equation}

On the other hand, a good approximation for CDM perturbations in this
epoch is \cite{Dodelson,Piattella:2018}
\begin{equation}
\delta_{CDM}=A\Phi_{\text{p}}\ln\left(Bx\right)=A\Phi_{\text{p}}\ln\left(\frac{B}{y_{\text{H}}}\right)+A\Phi_{\text{p}}\ln\left(y\right),\label{constraste CDM}
\end{equation}
where $x=y/y_{\text{H}}$, $A=9$, $B\simeq0.6237$.

Given the form of Eq.~(\ref{eq:appen-delta-m-rad}), we choose the
constants to be:
\begin{equation}
\bar{C}_{A}=A\Phi_{\text{p}}\frac{\sin\left(\alpha\ln\left(\frac{B}{y_{\text{H}}}\right)\right)}{\alpha}\text{ \ \ \ and \ \ \ \ }\bar{C}_{B}=A\Phi_{\text{p}}\cos\left(\alpha\ln\left(\frac{B}{y_{\text{H}}}\right)\right)\text{.}
\end{equation}
So we have the ansatz, (\ref{eq: warm-ansatz}),
\begin{equation}
\delta_{m}\left(y\right)\approx\frac{A\Phi_{\text{p}}}{\alpha}\sin\left[\alpha\ln\left(Bx\right)\right]\text{ , \ }x\gg 1\ \text{.}
\end{equation}

\section{Determination of $C_{A}$ and $C_{B}$}\label{app-C}

Now we address the issue of how to set the constants $C_{A}$ and
$C_{B}$, in Eq.~(\ref{eq:meszaros-final}). We impose that this solution
has to be compatible with the solution deep in the radiation era (\ref{eq: warm-ansatz-expansion}).
The solution is obtained by matching the functions and their first
derivatives at $y^{*}$ in the range $\left(10y_{H},10^{-1}\right)$.

Let's first expand $\delta_{A}$ and $\delta_{B}$ for $y\ll1$, we
also consider $\alpha^{2}\ll1.$ Then we get
\begin{equation}
\delta_{A}\left(y\right)\approx\frac{2}{3}\left(1-\frac{\alpha^{2}\ln^{2}y}{2}+\frac{3}{2}y\right).\label{delta A aprox 2}
\end{equation}
In order to obtain $\delta_{B}$, we substitute this expression for
$\delta_{A}$ in Eq.~(\ref{eq:delta-B-integral}), retaining only
terms linear in $y$ and quadratic in $\alpha$:
\begin{equation}
\delta_{B}\left(y\right)\approx\frac{2}{3}\left(1-\frac{\alpha^{2}\ln^{2}y}{2}+\frac{3}{2}y\right)\left[\int\frac{dy}{\frac{4}{9}\left(1-\frac{\alpha^{2}\ln^{2}y}{2}+\frac{3}{2}y\right)^{2}y\sqrt{1+y}}+C\right]\label{eq:delta-B-approx}
\end{equation}
The integration constant, $C$, is essential to guarantee that in
the limit $\alpha\rightarrow0$ we recover the known decaying solution
of the Mészáros solution. Solving the integral for $y\ll1$ and $\alpha^{2}\ll1$
we get
\begin{align}
\delta_{B}\left(y\right) & \approx\frac{3}{2}\left(1-\frac{\alpha^{2}\ln^{2}y}{2}+\frac{3}{2}y\right)\left[\int\left(1+\alpha^{2}\ln^{2}y-3y\right)\left(1-\frac{1}{2}y\right)\frac{dy}{y}+C\right]\Rightarrow\nonumber \\
\delta_{B}\left(y\right) & \approx\frac{3}{2}\left[\ln\left(\frac{y}{4e^{-3}}\right)+\frac{3}{2}y\ln\left(\frac{y}{4e^{-3}}\right)-\frac{7}{2}y-\frac{\alpha^{2}}{6}\left[\ln^{3}y-3\ln^{2}y\ln\left(4e^{-3}\right)\right]\right],\label{delta B aprox 2}
\end{align}
where $C=-\ln\left(4e^{-3}\right)$ ensures that the decaying solution
in the cold limit is recovered for $y\ll1$.

Finally, the general solution (\ref{eq:meszaros-final}) up to linear
order in $y$ and quadratic order in $\alpha$, is given by
\begin{multline}
\delta\left(y\right)\approx\frac{2}{3}C_{A}\left(1-\frac{\alpha^{2}\ln^{2}y}{2}+\frac{3}{2}y\right)\\
+\frac{3}{2}C_{B}\left[\ln\left(\frac{y}{4e^{-3}}\right)+\frac{3}{2}y\ln\left(\frac{y}{4e^{-3}}\right)-\frac{7}{2}y-\frac{\alpha^{2}}{6}\left[\ln^{3}y-3\ln^{2}y\ln\left(4e^{-3}\right)\right]\right].\label{eq:sol-gen-expan}
\end{multline}

Matching (\ref{eq:sol-gen-expan}), (\ref{eq: warm-ansatz-expansion})
and its derivatives at $y^{*}$, i.e., $\delta\left(y^{\ast}\right)=\delta_{m}\left(y^{\ast}\right)$
and $\partial_{y}\delta\left(y^{\ast}\right)=\partial_{y}\delta_{m}\left(y^{\ast}\right)$
yield the following system of equations
\begin{equation}
M\vec{C}=\vec{b}
\end{equation}
where
\begin{equation}
M=\left(\begin{array}{cc}
1-\frac{\alpha^{2}\ln^{2}y^{\ast}}{2}+\frac{3}{2}y^{\ast} & \ln\left(\frac{y^{\ast}}{4e^{-3}}\right)+\frac{3}{2}y^{\ast}\ln\left(\frac{y^{\ast}}{4e^{-3}}\right)-\frac{7}{2}y^{\ast}-\frac{\alpha^{2}}{6}\left[\ln^{3}y^{\ast}-3\ln^{2}y^{\ast}\ln\left(4e^{-3}\right)\right]\\
-\frac{\alpha^{2}\ln y^{\ast}}{y^{\ast}}+\frac{3}{2} & \frac{1}{y^{\ast}}+\frac{3}{2}\ln\left(\frac{y^{\ast}}{4e^{-3}}\right)-2-\frac{\alpha^{2}}{6}\left[\frac{3\ln^{2}y^{\ast}}{y^{\ast}}-6\frac{\ln y^{\ast}}{y^{\ast}}\ln\left(4e^{-3}\right)\right]
\end{array}\right)
\end{equation}
and
\begin{equation}
\vec{C}=\left(\begin{array}{c}
\frac{2}{3}C_{A}\\
\frac{3}{2}C_{B}
\end{array}\right)\text{ \ \ \ and \ \ \ }\vec{b}=\left(\begin{array}{c}
A\Phi_{\text{p}}\left[\ln\left(B\frac{y^{\ast}}{y_{\text{H}}}\right)-\frac{1}{6}\alpha^{2}\ln^{3}\left(B\frac{y^{\ast}}{y_{\text{H}}}\right)\right]\\
A\Phi_{\text{p}}\left[\frac{1}{y^{\ast}}-\frac{1}{2}\alpha^{2}\frac{\ln^{2}\left(B\frac{y^{\ast}}{y_{\text{H}}}\right)}{y^{\ast}}\right]
\end{array}\right).
\end{equation}
This system has a unique solution, $\vec{C}=M^{-1}\vec{b}$, given
by
\begin{align}
C_{A} & =\frac{3}{2}A\Phi_{\text{p}}\left\{ \ln\left(4e^{-3}\frac{B}{y_{\text{H}}}\right)-\frac{\alpha^{2}}{6}\ln^{2}\left(\frac{B}{y_{\text{H}}}\right)\left[\ln\left(\frac{B}{y_{\text{H}}}\right)-3\ln\left(\frac{e^{3}}{4}\right)\right]\right\} ,\label{CA outra possibilidade}\\
C_{B} & =\frac{2}{3}A\Phi_{\text{p}}\left\{ 1-\frac{\alpha^{2}}{2}\ln^{2}\left(\frac{B}{y_{\text{H}}}\right)\right\} .\label{CB}
\end{align}
Note that the solution does not depend on $y^{\ast}$, which indicates
that ansatz (\ref{eq: warm-ansatz-expansion}) and solution (\ref{eq:meszaros-final})
are equivalent in the range $10y_{\text{H}}\leq y\leq 10^{-1}$.

\bibliographystyle{JHEP.bst}
\bibliography{RefArqBibAbnt}

\end{document}